# First-Principles Mapping of the Electronic Properties of Two-Dimensional Materials for Strain-Tunable Nanoelectronics

Kostiantyn V. Sopiha[1,*], Oleksandr I. Malyi[2,*], Clas Persson[2]

[1]*Ångström Solar Center, Solid State Electronics, Department of Engineering Sciences, Uppsala University, Box 534, SE-75121 Uppsala, Sweden*
[2]*Centre for Materials Science and Nanotechnology, Department of Physics, University of Oslo, P. O. Box 1048 Blindern, NO-0316 Oslo, Norway*

***Abstract***

Herein, we demonstrate that first-principles calculations can be used for mapping electronic properties of two-dimensional (2d) materials with respect to non-uniform strain. By investigating four representative single-layer 2d compounds with different symmetries and bonding characters, namely 2d-$MoS_2$, phosphorene, α-Te, and β-Te, we reveal that such a mapping can be an effective guidance for advanced strain engineering and development of strain-tunable nanoelectronics devices, including transistors, sensors, and photodetectors. Thus, we show that α-Te and β-Te are considerably more elastic compared to the 2d compounds with strong chemical bonding. In case of β-Te, the mapping uncovers an existence of curious regimes where non-uniform deformations allow to achieve unique localization of band edges in momentum space that cannot be realized under either uniform or uniaxial deformations. For all other systems, the strain mapping is shown to provide deeper insight into the known trends of band gap modulation and direct-indirect transitions under strain. Hence, we prove that the standard way of analyzing selected strain directions is insufficient for some 2d systems, and a more general mapping strategy should be employed instead.

**Keywords:** non-uniform strain, band gap, 2d materials, spin-orbit coupling, strain engineering.

**\*E-mail:** kostiantyn.sopiha@gmail.com (K.V.S), oleksandrmalyi@gmail.com (O.I.M.)

## *1. Introduction*

Strain engineering has long been an important technique for governing electronic properties of solid-state materials. For instance, band gap tailoring through lattice mismatch in epitaxially grown semiconductor thin-films has now become a routine. However, despite its efficiency, a venue for strain engineering in epitaxial films is limited to static strains of only a few percent that can be introduced exclusively during the material deposition stage.[1] In these circumstances, atomically-thin two-dimensional (2d) materials emerge as perfect candidates for strain engineering as they are capable to withstand strains in the order of 10% without rupture or other irreversible changes of the structural integrity.[2] Another important advantage of 2d materials in the light of engineering is that the deformations can be dynamically modulated, which makes them attractive for application as strain-tunable transistors,[3] strain sensors,[4] and photodetectors.[5,6] As this material category already includes hundreds of synthesized and predicted 2d compounds,[7–9] a wide material selection is now available for functionalizing the strain-induced effects for advanced nanoelectronics devices.[3] In pursuit of these goals, great efforts have been made to uncover peculiarities of electronic structure changes in 2d materials with respect to strain. It has been revealed that strain-induced effects in 2d materials are strongly material-dependent.[10] For instance, tensile strain was shown to gradually reduce the band gap of most 2d materials but increase it in case of phosphorene (*i.e.*, single-layer black phosphorus).[1] While first-principles studies have already rationalized most of such phenomena, one key discrepancy between the experiment and theory persists. Specifically, experimentally, strain in 2d sheets can be applied in multiple directions through bending, folding, stretching, wrinkling, thermal expansion of the substrate, coherent epitaxial growth, and in principle, by various combination thereof.[1,2] In contrast, on the first-principles level, the deformations are mostly reduced to uniform and uniaxial strains, except several works introducing nonequal biaxial deformations in mutually perpendicular directions.[11,12] To bridge this gap, several authors have turned to mapping electronic structures of materials with respect to independent strain components.[13–15] Despite being naturally more informative, this approach has not gained due attention because its merits remain poorly outlined, and the number of considered 2d materials is still small.

Herein, we assess and adapt the methodology for mapping strain-induced changes in electronic structures of 2d materials for four representative compounds with different symmetries and bonding characters. Specifically, we consider 2d-$MoS_2$ monolayer as a classic example of 2d materials with strong chemical bonding; phosphorene as an example of highly anisotropic 2d compounds with covalent bonding; and two phases of newly-discovered 2d-Te[16] that have relatively low cohesive energies implying lower strength for the in-plane bonds. Within the focus on strain engineering, we (a) re-examine the strain dependence of electronic structures for the well-known 2d compounds, (b) provide a deep theoretical insight into the peculiarities of strain-induced effects in the weakly-bonded 2d-Te phases, and (c) discuss the role of symmetry in these phenomena. The obtained results are expected to advance the progress in functionalization of 2d materials through strain engineering not only for the four analyzed compounds but also for a wider family of 2d materials, in general.

## 2. Methods

All calculations presented in this study were carried out using Vienna *Ab initio* Simulation Package (VASP)[17–19] employing projector augmented wave (PAW) pseudopotentials[20,21] with valence electron configurations of $4d^5 5s^1$ for Mo, $3s^2 3p^4$ for S, $3s^2 3p^3$ for P, and $5s^2 5p^4$ for Te. The cut-off energy for plane-wave basis was set to 400 eV throughout the study. The ionic relaxations were performed using $\Gamma$-centered Monkhorst-Pack k-point grids[22] of 36×36×1, and 36×26×1, 28×28×1, 28×22×1 for primitive cells of 2d-$MoS_2$, phosphorene, α-Te, and β-Te, respectively. The relaxations were conducted until reaching 5 meV/Å convergence threshold for the atomic forces. To minimize spurious interaction between the periodic images, vacuum slab of at least 15 Å was introduced in the out-of-plane direction of all 2d systems. Since 2d-Te phases are characterized by the relatively weak in-plane Te-Te bonding, the van der Waals (vdW) contribution may play a more important role in the material stability, and thus, the computational results can be affected by non-local effects. In this regard, the long-range dispersion interactions were included employing optB86b-vdW functional[23,24] for all structure optimizations, including 2d-$MoS_2$ and phosphorene for consistency, unless otherwise specified. The energy-minimized stress-free primitive cells of all four considered materials, as optimized with optB86b-vdW functional, are given in Supporting Information, Structures S1-S4. To accurately predict electronic structures of 2d-Te systems, one self-consistent loop with spin-orbit coupling (SOC) was added after the ionic relaxations. This final loop was performed with Perdew-Burke-Ernzerhof (PBE) functional,[25] similarly to the approach employed elsewhere.[26–28] The k-point spacing was doubled (yielding twice spacer k-point grids) for the SOC calculations. Noteworthy, SOC was found to have a negligible effect on the geometries of all investigated systems. A detailed analysis of the impact of SOC and vdW forces on the computed properties is presented in Supporting Information, Figures S1-S4. The main results replicated using hybrid Heyd-Scuseria-Ernzerhof (HSE06) exchange-correlation functional[29] with the default mixing parameter of 0.25 are also given in Figure S5 for comparison.

Although some 2d materials can withstand deformations in the order of 10%,[30,31] in practice, strains of only a few percent are often used to avoid the material slippage over the substrate.[2,4,32,33] Intact with the experiments, our calculations were also constrained to deformations ranging from -5.0% to +5.0%. The strain maps for band gaps, positions of the conduction band minima (CBM), and positions of the valence band maxima (VBM) were constructed by deforming the equilibrium cells of 2d materials into two mutually perpendicular directions on the rectangular cells defined in Figure 1a by the black frames. As one can see, the X and Y axes are aligned with armchair and zigzag axes of the lattices whenever such definitions are applicable. After the ionic relaxations, the optimized 2d-$MoS_2$ and α-Te structures were converted to primitive cells (outlined by the blue frames in Figure 1a) for the band structure analysis. For each material, a total of 1681 systems were created by sampling the strain spectrum into uniform grids with 0.25% step. *Importantly, herein, we disregarded all shear deformations, which means that the obtained results are only applicable to a special case of biaxial strain with principal directions coinciding with axes of the defined coordinate systems.* Further, we do not consider the possibility of fracturing the 2d materials under tensile strain below +5.0%, which cannot be ruled out for phosphorene and 2d-Te due to the lack of experimental evidence. Moreover, the possibility of rippling under compressive strain,[34] which itself can affect the electronic properties of 2d-materials,[35] was not included. The deformation with equal values of strain along X and Y axes are referred to as "uniform" and called "non-uniform" otherwise. The positive and negative values of strain denote tension and compression, respectively. It should be noted that 2d-$MoS_2$ and α-Te reduce their layer group symmetry under non-uniform strain from hexagonal (layer group: $p$-6$m$2) and trigonal ($p$-3$m$1) to orthorhombic ($cm$2$m$) and monoclinic/orthogonal ($c$2/$m$11), respectively. In principle, such changes require redefining high symmetry points for band structure analysis. However, to avoid abrupt changes in the representation of band structures with respect to strain, herein, we instead duplicate some of the high symmetry points and adjust the k-paths every time when the changes in symmetry occur. The exact positions of high symmetry points for each lattice are illustrated in figures below.

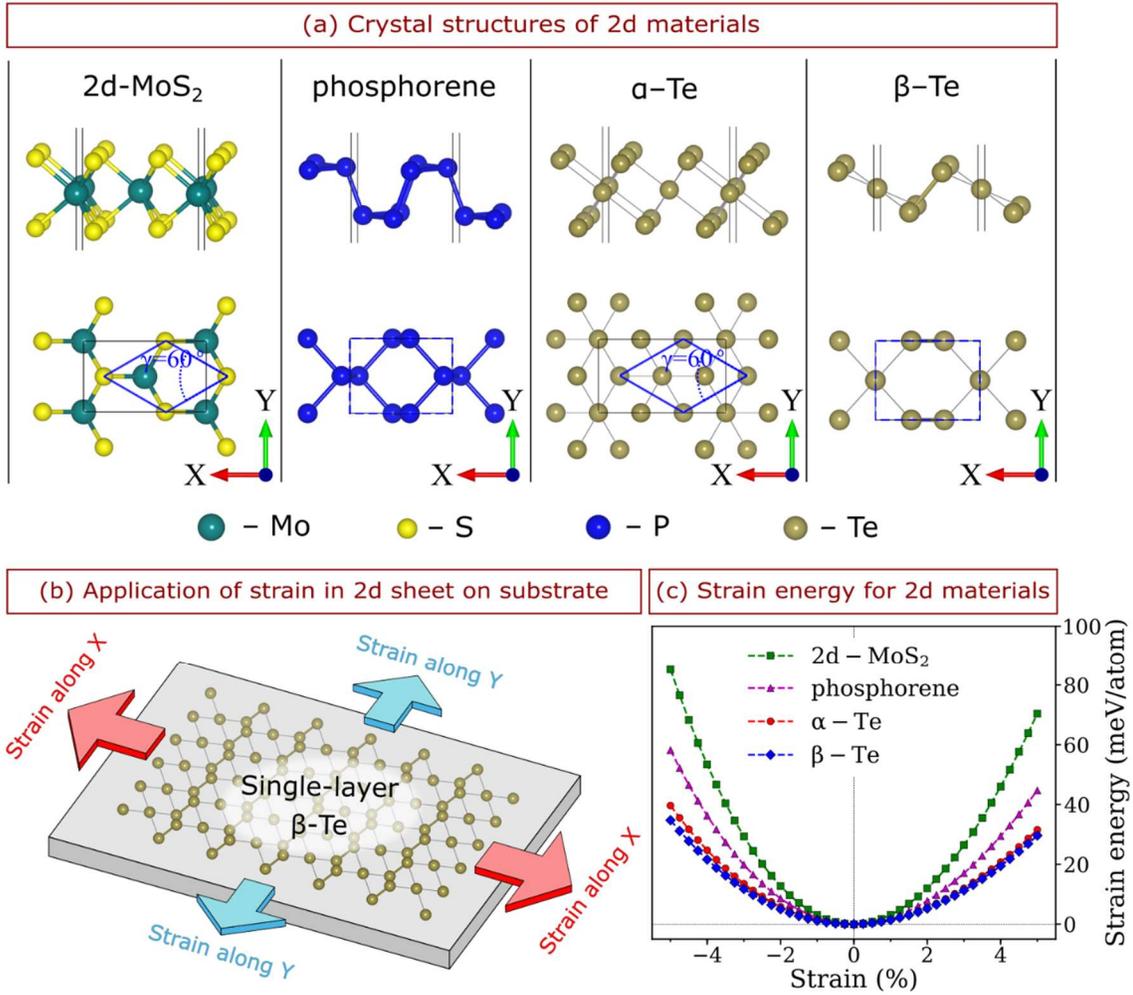

**Figure 1.** (a) Crystal structures of unstrained single-layer 2d-$MoS_2$, phosphorene, α-Te, and β-Te. The axes define directions of the applied strain, the black and the blue frames show boundaries of the rectangular and primitive cells, respectively. (b) Schematic illustration depicting application of strain in 2d materials on example of single-layer β-Te. (c) Computed dependencies of strain energies on uniform strain for these materials. The strain energy is calculated as a difference in energies of the undeformed and deformed systems.

### *3. Results and Discussion*

#### *3.1. On elasticity of the 2d materials*

Elasticity is a key parameter for strain-induced band gap engineering, which is often realized by deforming a flexible substrate with an adherent 2d material, as illustrated in Figure 1b. As such, not only the high elasticity promotes strain transfer from the substrate but also allows a higher degree of strain without slippage, thus widening the functional range of deformations.[2] Herein, elasticities of the 2d-materials were compared based on the computed dependences of total energy on applied strain (*i.e.*, strain energies). Since mechanical properties of β-Te and phosphorene are anisotropic,[36–38] as defined by their symmetries, the elasticity analysis was performed effectively by imposing the same strain along two mutually perpendicular in-plane directions (called "uniform" strain in this work). This rather crude approach allows a simple but qualitative comparison of the elasticities of 2d materials (more information on strain energies for the uniaxial deformations can be extracted from the strain maps shown in Figures S1-S4). As one can see, the strain energies induced in α-Te and β-Te are roughly half of that in 2d-MoS$_2$ and about 50% lower than that in phosphorene, irrespective of the magnitude of the applied strain (see Figure 1c). For the reference, the computed strain energies for 2d-MoS$_2$, phosphorene, α-Te, and β-Te undergoing +5.0% tensile strain are 70.4, 44.7, 31.6, and 29.6 meV/atom, respectively. These results illustrate that the bonds in 2d-MoS$_2$ and phosphorene are stiffer than those in 2d-Te, in agreement with notably low cohesive energies of the later.[16] Hence, both α-Te and β-Te are more elastic, which can open more doors for 2d-Te in strain-tunable nanoelectronics. As most of these applications utilize strain-induced modulations of the material properties,[1] the corresponding changes in electronic structures of the 2d materials are also analyzed and discussed in detail below.

#### *3.2. Strain-induced effects in 2d-MoS$_2$*

The undeformed single-layer 2d-MoS$_2$ has *p*-6*m*2 layer group symmetry that changes to *cm*2*m* upon non-uniform deformation. Experimentally, this change can be observed as a split in characteristic Raman peaks upon uniaxial[32,39,40] but not uniform[41] strain. These results thus highlight the importance of mapping electronic structures of 2d materials upon non-uniform deformations. The computed changes in the electronic structure of 2d-MoS$_2$ with respect to strain is presented in Figure 2. In agreement with previous studies,[42] our calculations show that undeformed 2d-MoS$_2$ has a direct band gap of 1.67 eV with the band edges at *K* point. The undeformed 2d-MoS$_2$ is also characterized by 0.15 eV spin-orbit splitting of the VBM orbitals at *K* point into spin-up and spin-down bands.[42–44] Experimentally, this band split is seen as two close-energy peaks in the absorption spectra observed for MX$_2$ (X = S, Se; M = Mo, W) compounds.[44–46] Notably, the band gap energy *versus* strain curves in Figure 2a obtained for the uniaxial strain modes (along X and along Y) nearly overlap, in agreement with previous studies.[12,32] Herein, we refer to the results for strain along X axis in 2d-MoS$_2$ as uniaxial, unless otherwise specified. Further, we determine the rate of band gap change as an absolute value of the slope of the band gap versus energy curve in the point of interest. For small regions from –0.75% to +1.25% uniform strain and from –1.25% to +2.5% uniaxial strain, the band gap decreases by 113 and 58 meV/%, respectively, in good agreement with experimental values of about 100 and 50 meV/%.[2,32,33,40,41] All band gap *versus* strain curves in Figure 2a have two distinct bending points that reflect abrupt changes in the electronic properties. The first such point at about +1.25% uniform strain and +2.5% uniaxial tensile strain marks a transition of the VBM to $\Gamma$ point, which is a known direct-to-indirect band gap transition in 2d-MoS$_2$.[12,32,33,40,47,48] Beyond the transition point, the rate of band gap change increases to about 207 and 110 meV/% for uniform and uniaxial strains, respectively, which are the largest values found in 2d-MoS$_2$ within the investigated strain range. Such a strong band gap modulation effect indicates that a much greater response of 2d-MoS$_2$ to all modes of strain can be engineered by pre-straining it uniformly more than +2.5% before the characterization. The second bending point reflects another direct-to-indirect band gap transition occurring under compression when the band gap reaches a maximum of 1.75 eV and CBM jumps to $\Gamma$-$K_3$ symmetry line at –0.5% uniform or –1.25% uniaxial compression, respectively. Upon further compression, the slope of the band gap curve changes sign, and the rate of band gap modulation becomes approximately 64 meV/% for uniform and 37 meV/% for uniaxial deformations. To the best of our knowledge, the second direct-to-indirect transition has only been predicted computationally[11,47] due to practical difficulties associated with applying large compression strains in atomically thin 2d sheets.[1] However, the existence of such a transition might be important to rationalize the observed trends in photoluminescence (PL) for MoS$_2$-WS$_2$ heterobilayer.[49]

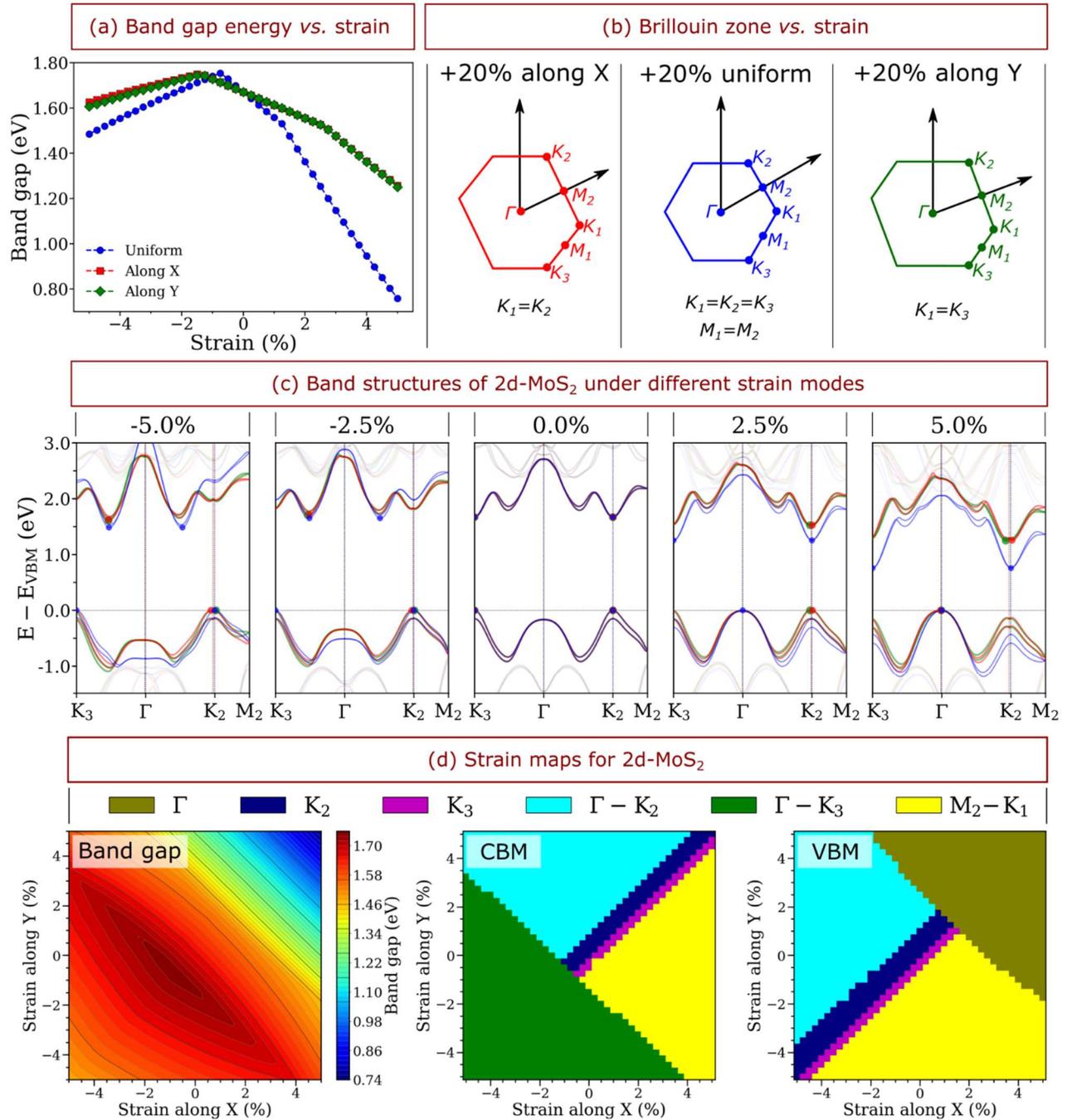

**Figure 2.** Electronic structures of single-layer 2d-MoS₂ under strain. (a) Band gap dependence on the applied strain. (b) Brillouin zones of the deformed MoS₂ (excessive +20% elongation is applied here for illustration purposes). (c) Band structures of 2d-MoS₂ under different strain modes; the red, blue, and green bands correspond to "uniform", "along X", and "along Y" modes, respectively; all except the highest occupied and lowest unoccupied bands are dimmed for clarity. (d) Strain maps of band gap, CBM, and VBM in momentum space. More details on the band structure variations in 2d-MoS₂ with respect to strain are given as Web Enhanced Object (Video 1).

Despite the apparent similarities in the band gap *versus* strain curves under the deformations along X and along Y, a close inspection of the results in Figure 2c,d reveals that the underlying processes have one key distinction. Specifically, when 2d-MoS$_2$ is elongated along X axis, both the CBM and VBM drift away from the $K$ point along $M_2$-$K_1$ symmetry line, whereas elongation along Y axis leads to a similar drift along $\Gamma$-$K_2$ line instead. Although such changes are likely insufficient to noticeably impact the electrical conductivity, they might be critical for the effects associated with valley polarization.[50] Importantly, the magnitudes of the VBM and CBM shifts in the momentum space under the strain are close but not equivalent, which means that the band gap of 2d-MoS$_2$ goes from direct to near-direct under small uniaxial deformation. Considering that near-direct nature of band gap may hinder PL, this change in the band structure may additionally contribute to the gradual decrease in PL intensity of 2d-MoS$_2$ under strain.[40]

### 3.3. Strain-induced effects in phosphorene

Unlike single-layer 2d-MoS$_2$, both undeformed and deformed phosphorene have orthorhombic (*pman*) layer group symmetry with distinct structural anisotropy that translates into anisotropy of mechanical, electrical, and optical properties.[37,38,51–55] As can be seen in Figure 3, unstrained phosphorene is a semiconductor with computed near-direct band gap of 0.79 eV. Here, the CBM and VBM are at $\Gamma$ point and $\Gamma$-$Y$ symmetry line near $\Gamma$ point, respectively, in good agreement with previous studies.[51,56–58] Notably, the direct energy gap at $\Gamma$ point is only several meV larger than the near-direct gap, implying that the result can be sensitive to the computational setup. Irrespective of the deformation mode analyzed, the band gap of phosphorene increases near-linearly under strain, except for a small region beyond reaching a maximum of 1.21 eV at +4.25% uniform tensile strain (see Figure 3a). In the broad linear regimes, the computed rates of band gap modulation under the uniform, along X, and along Y strains are 107, 51, and 59 meV/%, respectively. These values are roughly half of those measured for multilayer black phosphorus (99 and 109 meV/% along X and along Y axes, respectively).[4] Most likely, this discrepancy between the computational and experimental results reflects another form of thickness dependence of electronic properties in phosphorene,[4,52–54,59,60] as can be concluded from the comparison of band gap *versus* strain curves for monolayer and multilayer black phosphorus reported elsewhere.[13,52,61] Owing to the relatively small energy gap of phosphorene, this modulation manifests itself experimentally as piezoresistive effect.[4] Despite the structural anisotropy, both uniaxial strain modes yield similar dependencies of band gaps on the applied strain, in agreement with the experiment.[4] However, the anisotropy is evident from the position of the VBM point and curvature of the lowest conduction bands. First, while phosphorene deformed along X axis retains its near-direct band gap character, +0.75% uniform strain or +1.0% strain along Y axis makes the band gap direct by gradually shifting the VBM level to $\Gamma$ point. From the analysis of band structures in Figure 3c, however, it is likely that a similar drift of the VBM to $\Gamma$ point can occur under strain along X axis, although such transition would require elongations that lay outside the investigated strain region. Second, +4.25% uniform strain alters the band order at CBM, which abruptly changes the band curvature at CBM and leads to the appearance of bending point in the band gap *versus* strain curve (see Figure 3a). The impact of this phenomenon on anisotropy of the electron mobility has been discussed earlier.[51] Contrary, the same magnitude of strain along X maintains the conduction band curvature. Finally, by extrapolating the band structure changes beyond the boundary of investigated strain range, one could predict an upcoming transition of the CBM to $\Gamma$-$X$ symmetry line upon further elongation along Y axis.[57]

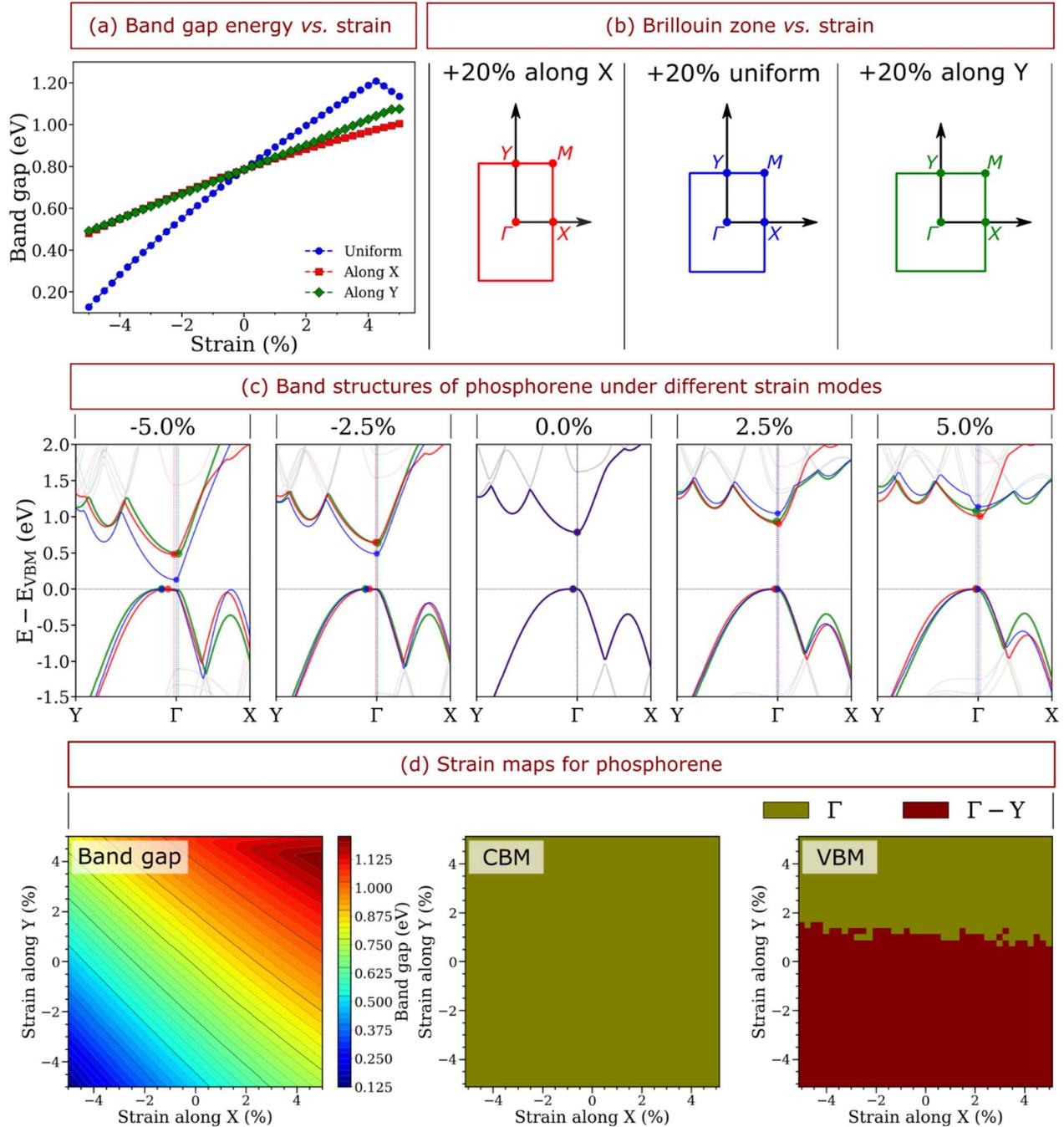

**Figure 3.** Electronic structures of single-layer phosphorene under strain. (a) Band gap dependence on the applied strain. (b) Brillouin zones of the deformed phosphorene (excessive +20% elongation is applied here for illustration purposes). (c) Band structures of phosphorene under different strain

modes; the red, blue, and green bands correspond to "uniform", "along X", and "along Y" modes, respectively; all except the highest occupied and lowest unoccupied bands are dimmed for clarity. (d) Strain maps of band gap, CBM, and VBM in momentum space. More details on the band structure variations in phosphorene with respect to strain are given as Web Enhanced Object (Video 2).

### 3.3. Strain-induced effects in 2d-Te

The possibility to synthesize two-dimensional polymorphs of tellurium was first explored using first-principles calculations.[16] The original work predicted the existence of two phases of 2d-Te with close energies (denoted α-Te and β-Te according to the original study, although alternative nomenclature coexists[62,63]). Herein, we find that the ground state phase of single-layer 2d-Te depends on the choice of exchange-correlation functional, with PBE and optB86b-vdW stabilizing β-Te and α-Te by a margin of 25 and 58 meV/atom, respectively. In the multilayer form, different 2d-Te phases were suggested to reconstruct from bulk-like Te film with so-called "magic" thicknesses.[16] Notably, unlike 2d-MoS$_2$ and phosphorene, the predicted 2d-Te phases do not have layered bulk counterparts. The experimentally grown multilayered flakes of Te resemble either the helical-chain structure of the bulk[64,65] or β-Te phase,[16] with the transition between them occurring at a certain thickness.[66] At the same time, the α phase of 2d-Te has only been observed as patches on rippled tri-layer Te.[67] From one side, the suppressed growth of α-Te might be due to a greater structural difference between the bulk and α-Te, which invokes higher barrier for reconstruction of the bulk-like film to α-Te than to β-Te. From the other side, the small energy difference may not provide enough thermodynamic stimulus to drive the reconstruction from the β-Te to α-Te phase. Moreover, the isolation energies for single-layers of both phases are known to exceed those for conventional 2d materials,[7] indicating a key role of the substrate to stabilize 2d-Te sheets, which can also affect the relative stabilities of the competing Te phases. The breakthrough in a controllable synthesis of single-layer 2d-Te might be on the way considering that high elasticity makes 2d-Te phases particularly susceptible to strain engineering. In these regards, while experimentalists refine the synthesis strategies,[62] first-principles studies explore the 2d-Te phases for thermoelectrics,[68] solar cells,[69] etc. Therefore, herein, we further widen the realm of applications for 2d-Te by analyzing the strain dependences of electronic structures for both predicted phases.

### 3.3.1. Investigation of α-Te

Figure 4 summarizes computed electronic structures of α-Te under strain. As one can see, unstrained α-Te has an indirect band gap of 0.41 eV with the CBM at $\Gamma$ point and the VBM at $\Gamma$-$M_2$ symmetry line. When strained uniformly, the band gap decreases near-linearly by about 28 meV/% throughout the investigated range, except the short region of compression exceeding -4.25%. It should be noted that this rate of the band gap modulation is roughly four times smaller than those in 2d-MoS$_2$ (113 meV/%) and phosphorene (107 meV/%). Here, the maximum band gap of 0.53 eV at -4.25% strain marks an abrupt transition of the VBM from $\Gamma$-$M_2$ symmetry line to $\Gamma$ point. Considering that the CBM of α-Te is at $\Gamma$ point irrespective of the strain direction and magnitude (see Figure 4d), this change in the VBM manifests indirect-to-direct band gap transition under compression, as it was discussed in earlier.[70] In contrast, uniaxial strain reduces the layer symmetry of α-Te from $p$-3$m$1 to $c$2/$m$11, which is reflected in the strain dependence of the electronic properties. As can be seen in Figure 4a, the overall shape of the band gap *versus* strain curves resembles parabola for both uniaxial deformation modes, with the energy gap reaching 0.44 eV at -3.75% and 0.42 eV at -2.0% compression along X and along Y, respectively. The rates of band gap modulation under the uniaxial strains are 14 meV/% along X axis and 11 meV/% along Y axis, with the exact value depending on the strain magnitude. The computed curve for non-uniform deformation along Y exhibits an abrupt change in inclination at the tensile strain of +3.0%, reflecting a transition of the VBM from $\Gamma$-$M_2$ to $\Gamma$-$K_2$ symmetry line. Noteworthy, as evident from the strain maps shown in Figure 4d, the VBM point can only be shifted to $\Gamma$-$K_2$ symmetry line when the tensile strain along Y axis is greater than that along X axis. These results highlight the need for analyzing multiple strain directions for better understanding and control of electronic properties in 2d system. Nonetheless, the relatively small rates of the band gap modulation leave little space for the strain-induced band gap engineering in α-Te system.

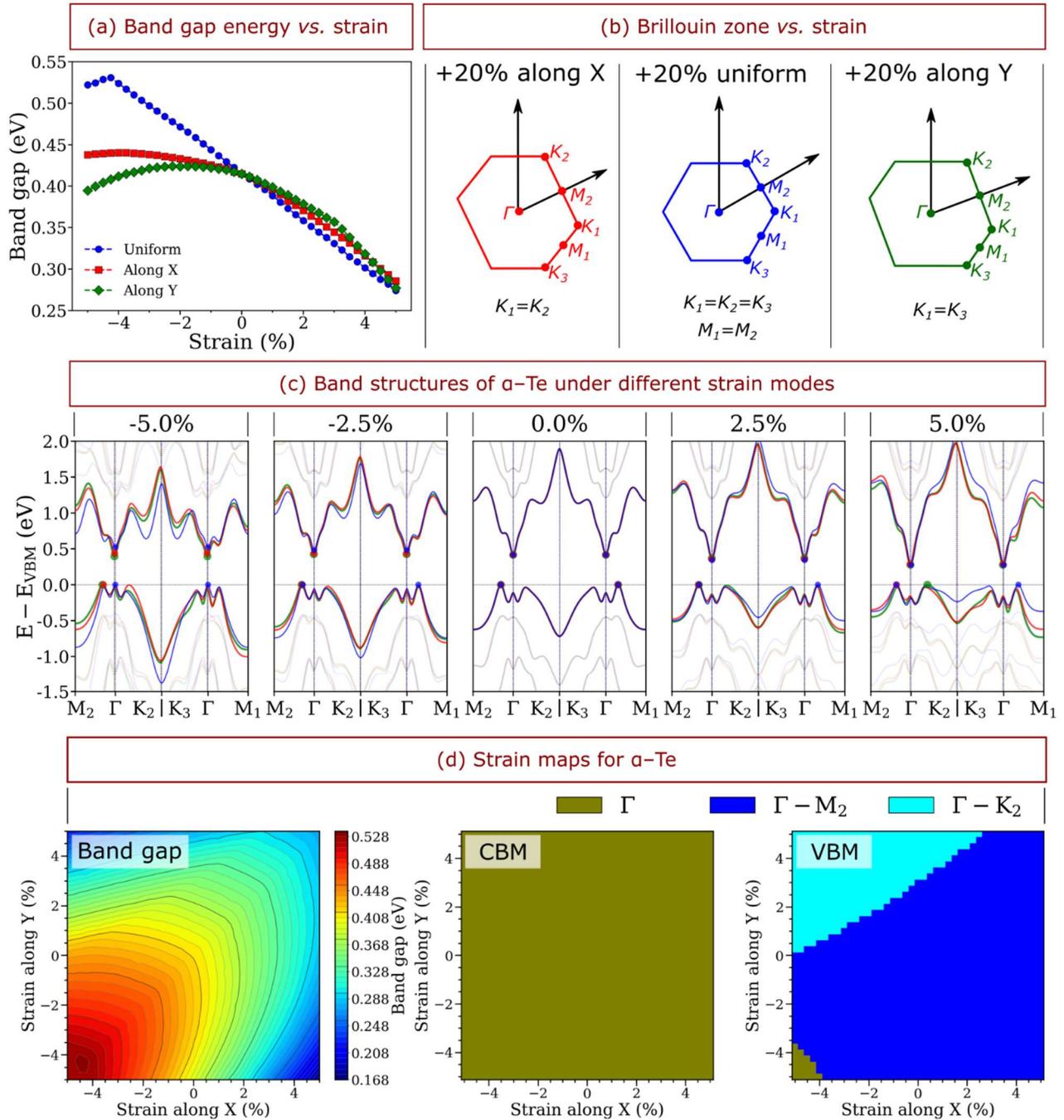

**Figure 4**. Electronic structures of single-layer α-Te under strain. (a) Band gap dependence on the applied strain. (b) Brillouin zones of the deformed α-Te (excessive +20% elongation is applied here for illustration purposes). (c) Band structures of α-Te under different strain modes; the red, blue, and green bands correspond to "uniform", "along X", and "along Y" modes, respectively; all except the highest occupied and lowest unoccupied bands are dimmed for clarity. (d) Strain maps of band gap, CBM, and VBM in momentum space. More details on the band structure variations in α-Te with respect to strain are given as Web Enhanced Object (Video 3).

### 3.3.2. Investigation of β-Te

The computed electronic properties of β-Te under strain are presented in Figure 5. As one can see, the undeformed β-Te is a direct band gap semiconductor with the computed energy gap of 1.02 eV and band edges at $\Gamma$ point, similarly to the results presented earlier.[16,70] The direct gap nature persists under tensile strain in either of the three analyzed deformation modes despite the energy gap decreases by 30 meV/% upon small elongation applied uniformly or along X axis (see Figure 5a). Noteworthy, β-Te retains the monoclinic/orthogonal ($p2/m11$) layer group symmetry irrespective of the applied strain. For deformations below ±1.0% along Y axis, either tensile or compressive, the band gap of β-Te stays roughly constant. When β-Te is compressed uniformly or along Y axis by -0.75%, it undergoes direct-to-indirect transition with the CBM point shifting to $X$ point, in consistence with the corresponding transformation reported before.[70] This transition gives rise to maxima points in the band gap *versus* strain curves where the gap reaches 1.03 eV for uniform strain and 1.02 eV for strain along Y axis, after which the band gap decreases by 99 and 93 meV/%, respectively. Noteworthy, the rate of band gap modulation in this region is the highest for 2d-Te phases and is comparable with those for single-layer 2d-MoS$_2$ (113 meV/%) and phosphorene (107 meV/%). This result suggests that the rate of band gap modulation in β-Te could be increased by pre-compressing it uniformly or along Y axis prior to the tests. Contrary, for β-Te deformed along X axis, the compression part of the band gap *versus* strain curve exhibits two key distinctions from those for the other modes. First, instead of decreasing upon compression, as seen under the uniform strain, the band gap energy keeps on increasing as β-Te is compressed along X axis until reaching a maximum of 1.08 eV at +4.25% strain. This behavior results in 0.46 eV difference between the energy gaps of β-Te compressed by +5.0% uniformly and along X axis. Second, the increase in band gap is accompanied by the CBM briefly shifting from $\Gamma$ point to $\Gamma$-$X$ symmetry line before finally transiting to $X$ point. This complex character of CBM transitions occurs because the lowest unoccupied band of β-Te compressed along X axis is relatively flat along $\Gamma$-$X$ symmetry line, with three competing band minima prevailing at different strain magnitudes, as shown in Figure 5c. Such a strong anisotropy of the band gap modulation under compression distinguishes β-Te from the other 2d materials discussed herein. While one could attribute this feature to the structural anisotropy, the real picture is more complicated as much weaker anisotropy of the band gap modulation was found in phosphorene. Hence, a detailed analysis of the correlation between the structural and the electronic modulation anisotropies is needed to pinpoint the exact origin of the effect in β-Te.

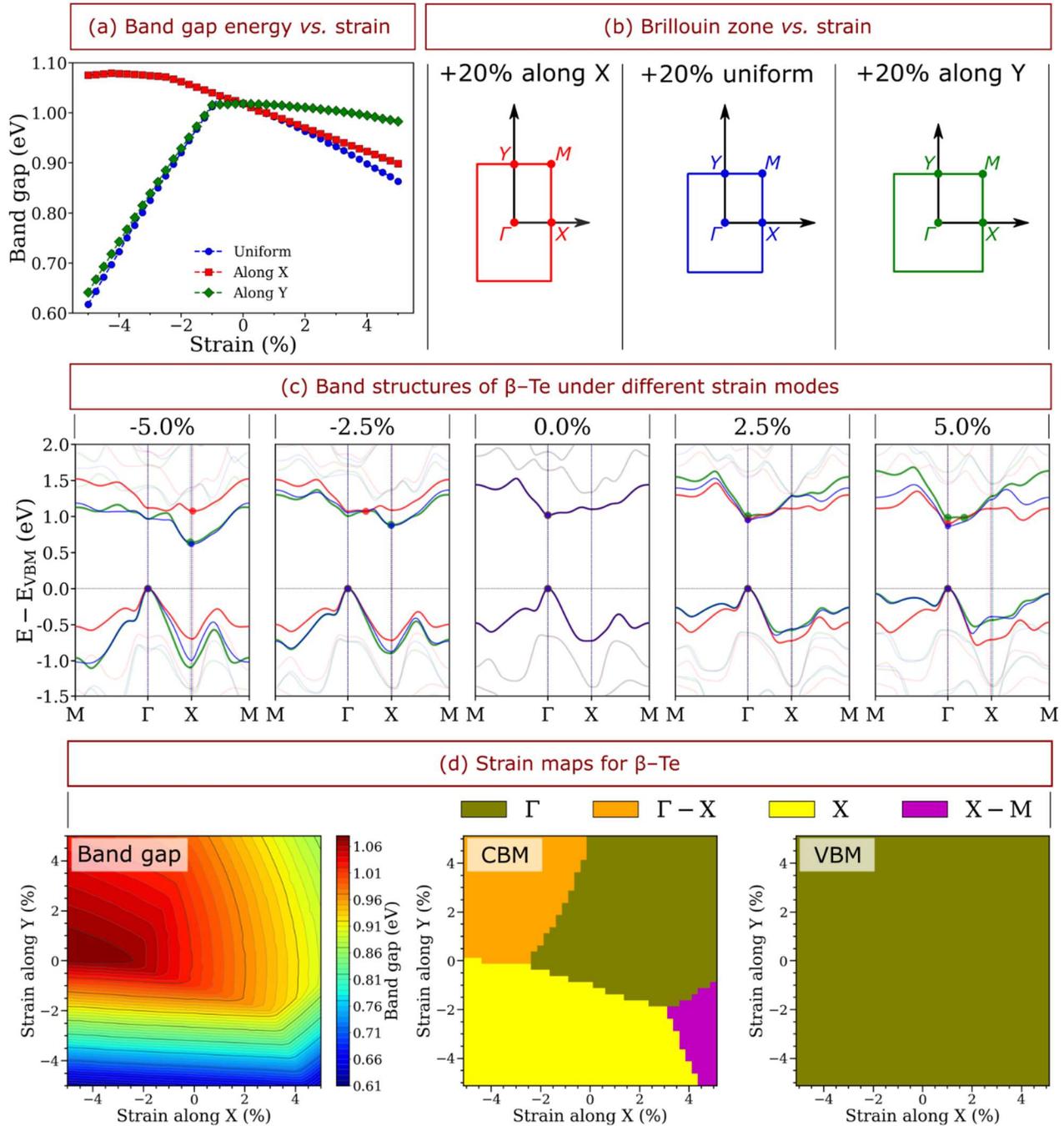

**Figure 5.** Electronic structures of single-layer β-Te under strain. (a) Band gap dependence on the applied strain. (b) Brillouin zones of the deformed β-Te (excessive +20% elongation is applied here for illustration purposes). (c) Band structures of β-Te under different strain modes; the red, blue, and green bands correspond to "uniform", "along X", and "along Y" modes, respectively; all except the highest occupied and lowest unoccupied bands are dimmed for clarity. (d) Strain maps of band gap, CBM, and VBM in momentum space. More details on the band structure variations in β-Te with respect to strain are given as Web Enhanced Object (Video 4).

Another distinct feature of β-Te is the complexity of strain-induced electronic structure changes, which is well illustrated by the strain maps shown in Figure 5d. As one can see, the CBM map for β-Te is shaped by four distinct regions, namely $\Gamma$, $X$, $\Gamma$-$X$, and $X$-$M$. Notably, the $X$-$M$ region cannot be observed under either of the three analyzed deformation modes, while the $\Gamma$-$X$ region can only be briefly entered by compressing β-Te along X axis. Instead, these two regions represent β-Te compressed in one direction while stretched in the other. Specifically, the $\Gamma$-$X$ section of the CBM map describes β-Te stretched along Y axis and compressed along X axis (top-left corner in the CBM map), whereas the $X$-$M$ region represents the material undergoing elongation along X axis and compression along Y axis (bottom-right corner in the CBM map). Since such non-uniform deformations are not usually explored in controllable single-flake experiments, these regions of electronic properties could easily be missed out unless specifically pointed out by the strain maps. This conclusion confirms that the strain mapping is indeed a very powerful tool for strain engineering in 2d materials. At the same time, the unique anisotropy of β-Te provides an opportunity to develop devices sensitive to both the magnitude and direction of the applied strain. However, at the moment, more research is needed to establish technical aspects for implementing these effects into practical devices.

### *Conclusions*

Owing to the ability of withstanding large tensile strains without rapture, 2d materials have emerged as ideal candidates for advanced strain engineering. Furthermore, the possibility to modulate dynamically their electronic properties by strain enables the application of 2d materials in strain-tunable nanoelectronics, including transistors, sensors, and photodetectors. Herein, by investigating four representative 2d materials with different symmetries and bonding characters (2d-$MoS_2$, phosphorene, α-Te, and β-Te), we demonstrate that the electronic structure changes occurring under strain are highly material- and strain-dependent. The differences in the strain effects are especially well illustrated *via* the construction of strain maps for some fundamental electronic properties. Apart from explaining the general features of band gap modulation, the strain mapping is proven to be important for marking transitions of the CBM and VBM levels in the momentum space that can lead to abrupt changes in the electronic properties. While the deformations along selected directions are generally sufficient to determine the order of changes, it cannot evince that all peculiarities of the underlaying processes are captured. As an example, by using strain mapping, we identified one region of non-uniform strain for β-Te where the CBM is at $X$-$M$ point, which cannot be achieved by either uniform or uniaxial deformations. Interestingly, pre-compressed β-Te exhibits a high rate of the band gap modulation that is closely comparable with that for 2d materials with strong chemical bonding. Furthermore, owing to the structural anisotropy, β-Te is distinguished by strong anisotropy of the strain-induced electronic changes. As a result, the constructed strain maps uncover a complex order of CBM transitions, which indicate that β-Te can indeed be attractive for the strain engineering purposes. At the same time, by exploring the electronic changes in other materials, we conclude that the strain maps can provide an additional advantage of predicting the CBM and VBM transitions under all possible straining routes, thus establishing it as a highly beneficial complementary technique for rationalizing the electronic structure changes in deformed 2d materials, in general.

### Supporting Information

Figures showing strain maps for strain energy and electronic structures of single-layer 2d-$MoS_2$, phosphorene, $\alpha$-Te, and $\beta$-Te constructed using different computational setups; Crystallographic Information File (CIF) structures for the energy-minimized stress-free primitive cells optimized using optB86b-vdW functional.

### Web Enhanced Objects

Videos summarizing evolution of electronic properties of 2d-$MoS_2$, phosphorene, $\alpha$-Te, and $\beta$-Te with respect to strain applied along selected deformation modes; these videos are extensions of Figures 2-5.

### Acknowledgments

This work is financially supported by the Research Council of Norway (ToppForsk project: 251131). We acknowledge Norwegian Metacenter for Computational Science (NOTUR) and Swedish National Infrastructure for Computing (SNIC) for providing access to the supercomputer resources.

Supporting Information for:

# First-Principles Mapping of the Electronic Properties of Two-Dimensional Materials for Strain-Tunable Nanoelectronics

Kostiantyn V. Sopiha[1,*], Oleksandr I. Malyi[2,*], Clas Persson[2]

*[1]Ångström Solar Center, Solid State Electronics, Department of Engineering Sciences, Uppsala University, Box 534, SE-75121 Uppsala, Sweden*
*[2]Centre for Materials Science and Nanotechnology, Department of Physics, University of Oslo, P. O. Box 1048 Blindern, NO-0316 Oslo, Norway*

**\*E-mail:** kostiantyn.sopiha@gmail.com (K.V.S), oleksandrmalyi@gmail.com (O.I.M.)

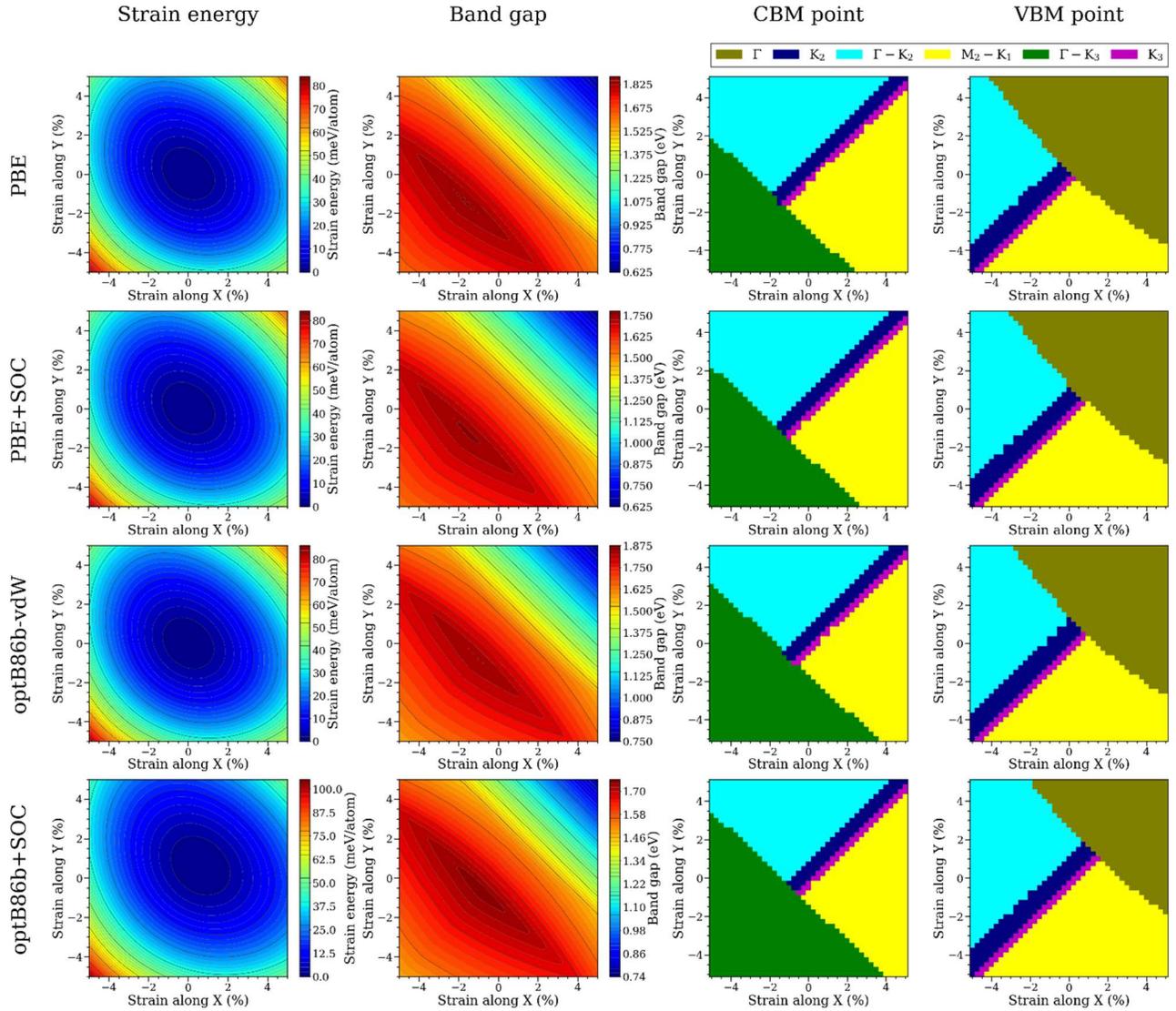

**Figure S1.** Strain maps for strain energy and electronic structures of single-layer 2d-MoS$_2$ constructed using different computational setups, from top to bottom: PBE; PBE with spin-orbit coupling (SOC) added in the last self-consistent loop; optB86b-vdW; PBE with SOC added in the last self-consistent loop for the geometries optimized with optB86b-vdW functional (these are the results presented in the main text).

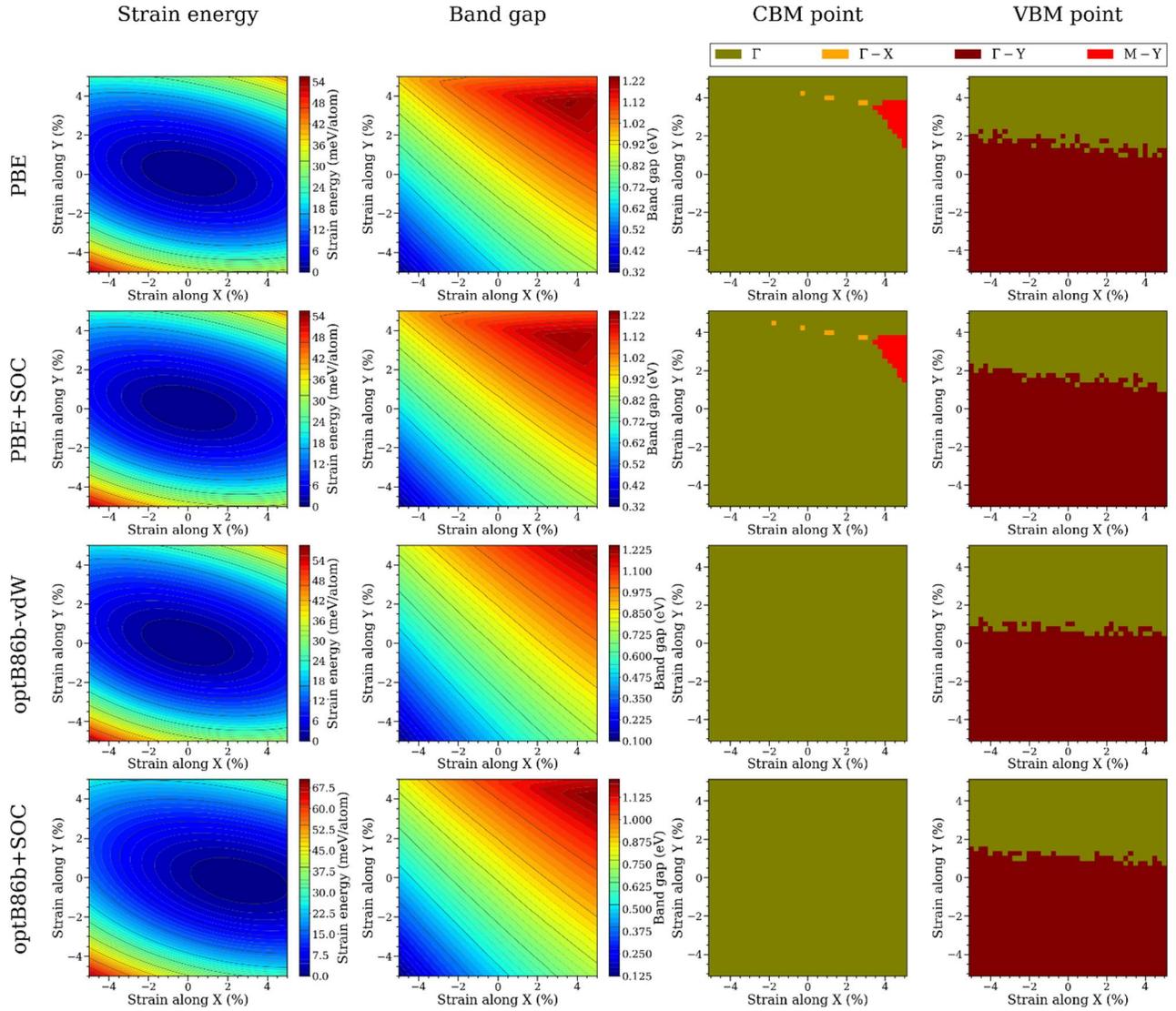

**Figure S2.** Strain maps for strain energy and electronic structures of single-layer phosphorene constructed using different computational setups.

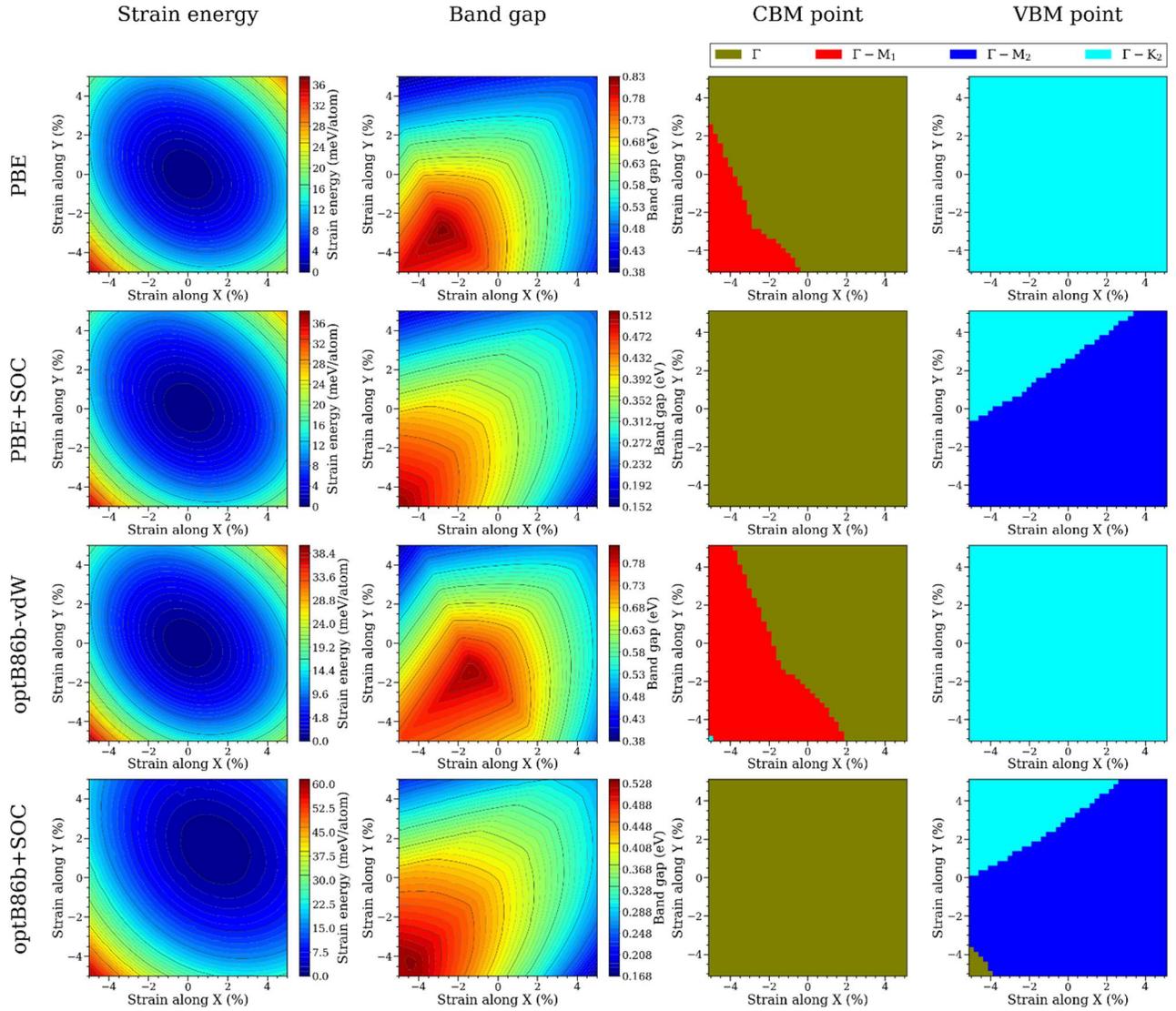

**Figure S3.** Strain maps for strain energy and electronic structures of single-layer α-Te constructed using different computational setups.

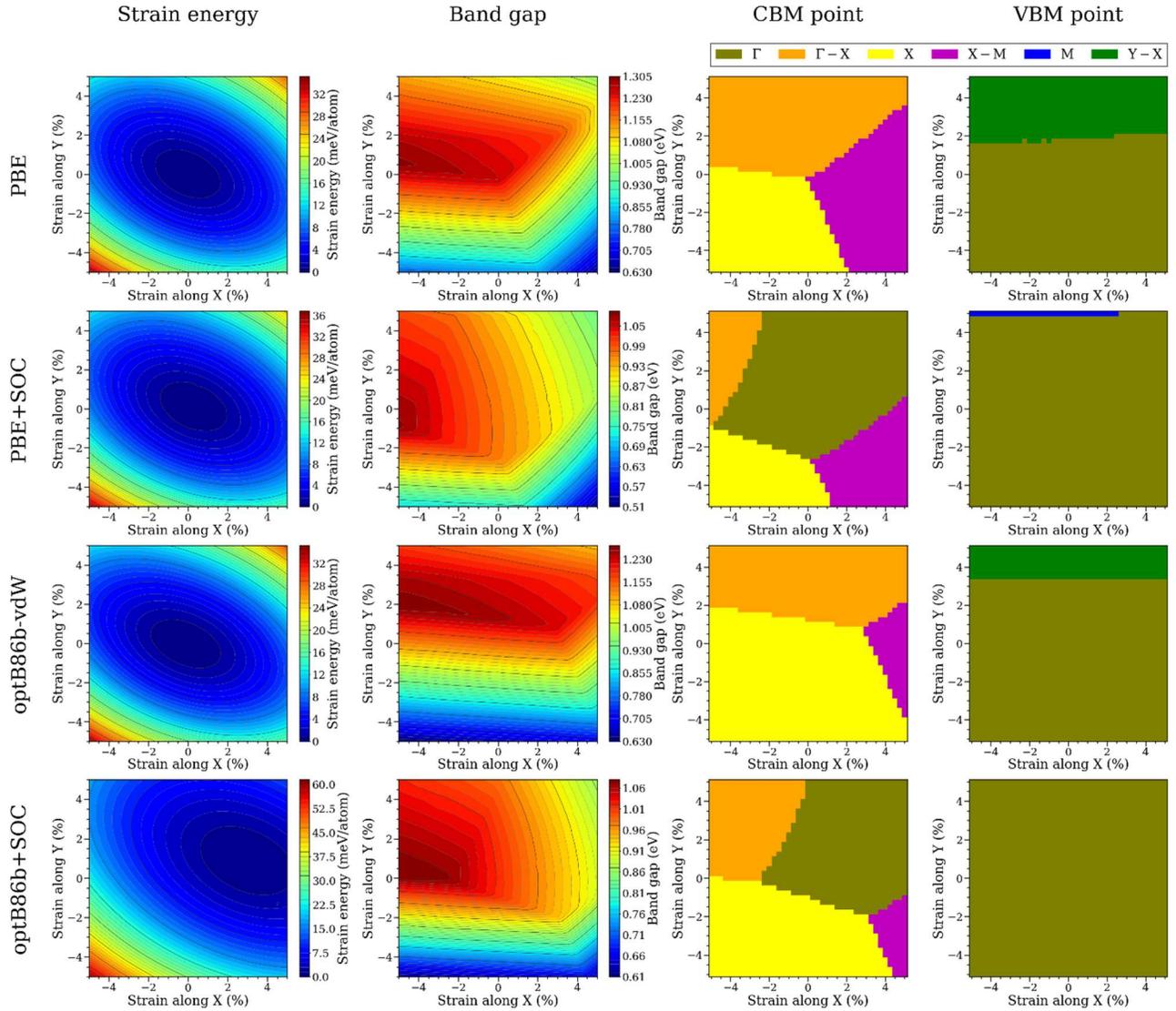

**Figure S4.** Strain maps for strain energy and electronic structures of single-layer β-Te constructed using different computational setups.

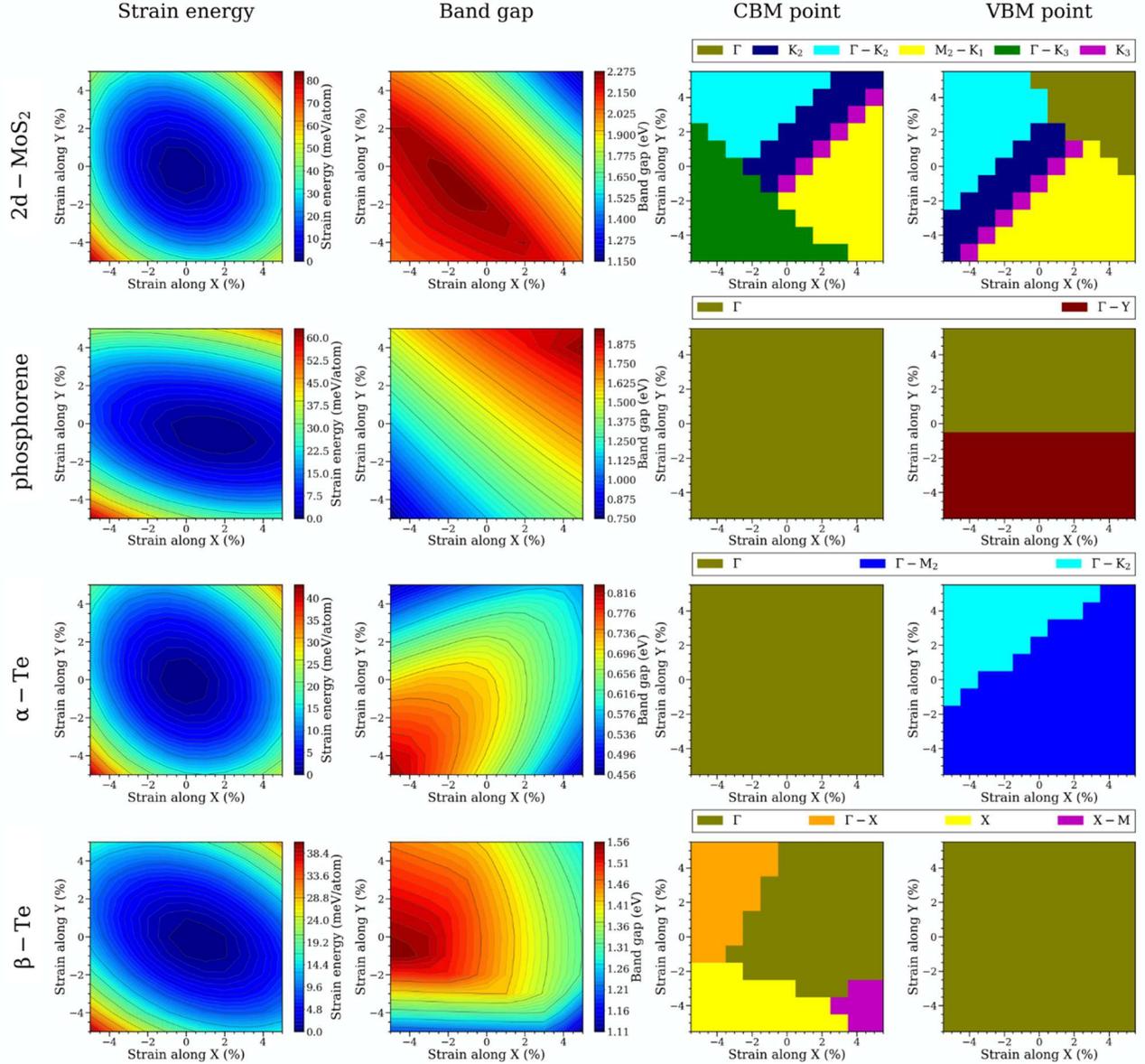

**Figure S5.** Strain maps for strain energy and electronic structures of single-layer 2d-MoS$_2$, phosphorene, α-Te, and β-Te constructed using HSE06 functional with inclusion of SOC. The grid resolution for the maps was reduced from 41×41 to 11×11 points. The properties are extracted after one self-consistent loop with HSE06 functional on the structures optimized using optB86b-vdW setup. No other computational parameters were changes compared to the optB86b-vdW setup.

**Structure S1.** Energy-minimized stress-free primitive cell of 2d-MoS$_2$ optimized using optB86b-vdW functional (in Crystallographic Information File (CIF) format).

```
_cell_length_a              3.16418
_cell_length_b              3.16418
_cell_length_c              18.34881
_cell_angle_alpha           90
_cell_angle_beta            90
_cell_angle_gamma           120
loop_
  _atom_site_fract_x
  _atom_site_fract_y
  _atom_site_fract_z
  _atom_site_type_symbol
  0.000000    0.000000    0.500000    Mo
  0.666667    0.333333    0.585636    S
  0.666667    0.333333    0.414364    S
```

**Structure S2.** Energy-minimized stress-free primitive cell of phosphorene optimized using optB86b-vdW functional (in CIF format).

```
_cell_length_a          3.30451
_cell_length_b          4.50844
_cell_length_c          17.51992
_cell_angle_alpha       90
_cell_angle_beta        90
_cell_angle_gamma       90
loop_
  _atom_site_fract_x
  _atom_site_fract_y
  _atom_site_fract_z
  _atom_site_type_symbol
  0.500000    0.084506    0.439293    P
  0.500000    0.915494    0.560707    P
  0.000000    0.584506    0.560707    P
  0.000000    0.415494    0.439293    P
```

**Structure S3.** Energy-minimized stress-free primitive cell of α-Te optimized using optB86b-vdW functional (in CIF format).

```
_cell_length_a          4.18319
_cell_length_b          4.18319
_cell_length_c          24.27020
_cell_angle_alpha       90
_cell_angle_beta        90
_cell_angle_gamma       120
loop_
  _atom_site_fract_x
  _atom_site_fract_y
  _atom_site_fract_z
  _atom_site_type_symbol
 0.000000    0.000000    0.500000    Te
 0.333333    0.666667    0.575777    Te
 0.666667    0.333333    0.424223    Te
```

**Structure S4.** Energy-minimized stress-free primitive cell of β-Te optimized using optB86b-vdW functional (in CIF format).

```
_cell_length_a              4.19397
_cell_length_b              5.54296
_cell_length_c              22.92676
_cell_angle_alpha           90
_cell_angle_beta            90
_cell_angle_gamma           90
loop_
  _atom_site_fract_x
  _atom_site_fract_y
  _atom_site_fract_z
  _atom_site_type_symbol
0.500000    0.000000    0.500000    Te
0.000000    0.657151    0.546837    Te
0.000000    0.342849    0.453163    Te
```